\documentstyle{article}

\input pz.sty
\input epsf.sty

\begin{document}

\PZhead{9}{27}{2008}{14 February}

\PZtitletl{Photometric observations of two type II-P Supernovae:}{normal 
SN II-P 2004A and unusual SN 2004\lowercase{ek}}

\PZauth{D. Yu. Tsvetkov}
\PZinst{Sternberg Astronomical Institute, University Ave.13,
119992 Moscow, Russia; e-mail: tsvetkov@sai.msu.su}

\begin{abstract}
CCD $BVRI$ photometry is presented for type II Supernovae
2004A and 2004ek. SN 2004A is found to be a typical SN II-P,
with the shape of the light and color curves and maximum
luminosity closely matching those for SN 1999em. SN 2004ek 
shows unusual light curves with long flat plateau in the $B$ band,
two peaks in the $V$ and prominent brightening in the $R$ and $I$
bands, starting about 45 days past outburst. The brightness
decline after the plateau stage is probably quite slow. The plateau 
luminosity
is about 1.5 mag brighter than average for SN II-P.   
\end{abstract}

SN 2004A was discovered by Itagaki on January 9.84 UT at magnitude
15.7 (Nakano and Itagaki, 2004). It is located at 
$\alpha =16\hr43\mm01\sec.90, \delta =+36\deg 50\arcm 12\arcs.5$
(equinox 2000.0), which is approximately $22\arcs$ west and $17\arcs$
north from the poorly defined nucleus of Sc galaxy NGC 6207.
Kawakita et al. (2004) reported that spectra obtained on January 11
showed blue continuum and prominent H Balmer lines with P Cyg profiles,
indicating young SN II. The expansion velocity was estimates as 
12000 km s$^{-1}$.
The detailed photometric and spectroscopic study of SN 2004A was
presented by Hendry et al. (2006), they also reported probable 
identification of a progenitor star in pre-explosion $HST$ images.

\medskip

SN 2004ek was discovered by Boles, Puckett and Cox (2004) on September
9.97 UT in Sb galaxy UGC 724 at magnitude 17.1. The position of SN is 
$\alpha =1\hr09\mm58\sec.51, \delta =+32\deg 22\arcm 47\arcs.7$
(equinox 2000.0), which is $11\arcs$ west and $41\arcs$ north from the 
center of the galaxy. 
Modjaz et al. (2004) reported that spectrum obtained on September 14.47
was similar to the early spectra of SN 1993J and 
showed featureless blue continuum indicating young SN II. 
Filippenko et al. (2004) confirmed these conclusions with spectroscopic
observations carried out on September 24. The spectrum revealed blue
continuum with weak hydrogen Balmer and Fe II absorption lines, although
the weak H$\alpha$ line was almost entirely in emission. 

\medskip

We carried out observations of SN 2004A and 2004ek with the following
telescopes and CCD cameras: 60-cm reflector of 
Crimean Observatory of Sternberg Astronomical Institute (C60) 
equipped with Roper Scientific VersArray1300B CCD camera (a), 
or Apogee AP-7p camera (b); 50/70-cm meniscus telescope of
Crimean Observatory (C50) with Meade Pictor 416XT camera; 70-cm 
reflector in Moscow (M70) with Apogee AP-47p (a) or AP-7p (b)
cameras. On January 30 images of SN 2004A were obtained with
50-cm reflector at Tatranska Lomnica in Slovakia equipped 
with SBIG ST-10 camera (S50). 

The color terms for C60 and M70
were reported by Tsvetkov et al. (2006). The observations at C50
were carried out only with $V$ filter which
was close to standard system, and no correction was applied.
The $UBVRI$ passbands at S50 were quite close to standard, and
only minor corrections were needed. 

The standard image reductions and photometry were made using IRAF.\PZfm 
\PZfoot{IRAF is distributed by the National Optical Astronomy Observatory,
which is operated by AURA under cooperative agreement with the
National Science Foundation}
Photometric measurements of SNe were made relative to 
local standard stars using PSF-fitting with IRAF DAOPHOT package. 
Both SNe occurred quite far from the centers of their parent
galaxies. At the position of SN 2004ek practically no host
galaxy background was present, while for SN 2004A it was noticeable
on the frames taken with longer exposures, when SN was faint. 
Subtraction of host galaxy images was applied to some SN 2004A frames,
but the results were found practically identical
to those obtained without subtraction.
 
The magnitudes of local standards were calibrated on photometric
nights, when photometric standards were observed at different 
airmasses. They are reported in Table 1, the images of SNe with 
marked local standards are shown in Figs. 1,2.         
The results of photometry of supernovae are presented in 
Tables 2,3.

\medskip

{\bf SN 2004A}. We started monitoring this SN on 2004 January 30, 
21 days after discovery, and continued observations until November 21. 
The magnitudes for our local standards were 
derived also by Hendry et al. (2006), and the mean differences 
between the two data sets are:
$\overline{\Delta B}= 0.13 \pm 0.01; \overline{\Delta V}= 0.08 \pm 0.01;
\overline{\Delta R}=0.05 \pm 0.02; \overline{\Delta I}=0.05 \pm 0.02$.
The differences of $B$ and $V$ magnitudes are quite significant.
Hendry et al. (2006) presented also another set of magnitudes, based on 
data from SDSS DR4. We found that these magnitudes are in much
better agreement with our calibration: 
$\overline{\Delta B}=-0.02 \pm 0.03; \overline{\Delta V}=-0.04 \pm 0.02;
\overline{\Delta R}=-0.03 \pm 0.03; \overline{\Delta I}=0.04 \pm 0.01$.
However, Hendry et al. (2006) preferred their own calibration for
reduction of SN 2004A photometry. We corrected 
the magnitudes from Hendry et al. (2006) for the differences we found 
between the calibrations of local standards and plotted them 
together with our data in Fig. 3. Also shown are the results of
CCD photometry (mostly unfiltered) reported at SNWeb.\PZfm
\PZfoot{http://www.astrosurf.com/snweb2/2004/04A\_/04A\_Meas.htm}
They show large systematic differences compared to our data, but are 
tracing the early rising part of the light curve and so are
valuable for defining the explosion date. 

We fit the light curves of a typical SN II-P 1999em (Leonard et al., 2002;
Elmhamdi et al., 2003a; Hamuy et al., 2001) to the data for SN 2004A
and find a very good match, which is evident from Fig. 3. 
The comparison confirms that explosion occurred on 2004 January 6
(JD 2453011), as estimated by Hendry et al. (2006). The plateau
stage starts on JD 2453018. The end of plateau can be defined
in different ways; if we assume that it is marked by the point 
on the light curve, 
corresponding to half of the brightness drop from plateau to 
the linear tail, which is at about JD 242453124, then the
plateau lasted for 106 days. 
The maximum brightness was reached at about JD 2453075 in all bands
except $B$: $V_{max}=15.5, R_{max}=15.0, I_{max}=14.7$.
The rate of decline in $B$ at the plateau phase was 0.013 mag day$^{-1}$.
The linear fits 
to the $V$ and $R$ magnitudes after
JD 2453200 give the following decline rates:
$0.0099\pm 0.0009$ mag day$^{-1}$ in the $V$
and $0.0088\pm 0.0004$ mag day$^{-1}$ in the $R$ band. These rates
are very close to the decay slope of $^{56}$Co, which is 
0.0098 mag day$^{-1}$.

The color curves for SN 2004A are presented in Fig. 4 and are 
compared to the 
color curves of SN 1999em. The agreement is quite good, except some
difference of $(B-V)$ curves at JD 2453080-120. We have also 
obtained one estimate of $U$ magnitude, which is not included into
Table 2: on JD 2453034.70 $U = 16.29\pm 0.08$, corresponding to the 
color $(U-B)=0.14\pm 0.09$. This value is about 0.4 mag bluer than
the color of SN 1999em at comparable epoch. Still we conclude that
the color evolution of the two SNe is similar and the extinction for
them is also quite close. The Galactic extinction in the direction
of NGC 6207 is small: $A_V=0.05$ according to Schlegel et al. (1998).
For SN 1999em total extinction of $A_V\approx 0.3$ is accepted in
most of the works (Elmhamdi et al., 2003a). We conclude that for 
SN 2004A the total extinction is close to this value.  
 
The distance to NGC 6207 is rather controversial: the radial
velocity corrected for the Virgo infall is 1240 km s$^{-1}$ according 
to NASA/IPAC Extragalactic Database\PZfm
\PZfoot{http://nedwww.ipac.caltech.edu} (NED),
corresponding to distance of 17 Mpc with 
$H_0$=73 km s$^{-1}$ Mpc$^{-1}$. However, Hendry et al. (2006)
estimate the distance using $HST$ photometry of the brightest supergiants
and find out that this method gives distances in the 
range from 17.7 to 26.8 Mpc. Finally they accept the mean of 
different estimates, including kinematic, which is $20.3\pm 3.4$ Mpc. 
Recently the distance to NGC 6207 was also estimated by $I$-band
Tully-Fisher relation (Springob et al., 2007) and was found to 
correspond to radial velocity 1346 km~s$^{-1}$. This means the
distance of 18.4 Mpc, and we accept this value as more probable.
Using this distance and extinction $A_V=0.3$ we construct the
absolute $V$ light curve, which is illustrated in Fig. 7. 
The light curve of SN 1999em is plotted for comparison (with
Cepheid distance 11.7 Mpc and $A_V=0.3$).
Both SNe appear quite similar, although SN 1999em is about 
0.5 mag brighter at the plateau stage. 
With absolute magnitude at plateau of $M_V=-16.1$ SN 2004A  
certainly belongs to SN II-P with normal luminosity.

We can try to estimate also the amount of $^{56}$Ni ejected
in the explosion by three methods: comparison with the tail 
luminosity of SN 1987A; correlations between $^{56}$Ni
mass and plateau $M_V$ and the steepness parameter $S$, which is
the maximum gradient during the transition from plateau to the
tail, in mag day$^{-1}$
(Elmhamdi et al., 2003b). According to our data, the luminosity at
radioactive tail for SN 2004A is about 0.8 mag less than for
SN 1987A, and assuming that the same relation holds for bolometric
luminosity, we derive $^{56}$Ni mass in SN 2004A to be about 2.07
times less then in SN 1987A, that is 0.036 $M_{\odot}$. We estimate
the parameter $S$ for SN 2004A to be approximately 0.12 mag day$^{-1}$,
although the coverage of the light curve is not quite good at that
stage, this value corresponds to $M(^{56}{\rm Ni})=0.027 M_{\odot}$.
And finally, the  correlation between plateau luminosity and 
$^{56}$Ni mass yields $M(^{56}{\rm Ni})=0.039 M_{\odot}$. We conclude
that all estimates agree reasonably well and confirm that SN 2004A
is a normal SN II-P.

\medskip

{\bf SN 2004ek.} We observed this SN from 2004 September 14 (5 days
after discovery) until 2005 Ferbuary 10. The light curves are presented
in Fig. 5, where we also plotted unfiltered CCD magnitudes reported
at SNWeb.\PZfm
\PZfoot{http://www.astrosurf.com/snweb2/2004/04ek\_/04ek\_Meas.htm}
These data show the early rise of brightness and allow to 
conclude that the explosion likely occurred at about September 8 
(JD 2453256). SN 2004ek certainly belongs to the class of SN II-P, 
but the differences between the light curves of SN 2004ek and those for
typical SNe II-P is 
evident. In the $B$ band SN 2004ek has a long period 
(JD 2453290-340)
of constant brightness, while all normal SNe II-P have nearly
linear decline at that stage. In the $V$ there are two peaks on 
the light curve, one immediately after outburst and the second at 
about JD 2453330. In the $R$ and $I$ bands after the period
of constant brightness there is a prominent increase of luminosity,
amounting to 0.3 mag in $R$ and 0.4 mag in $I$. The early decline 
after the plateau is
probably quite slow, the gradient in $R$ is only 0.013 mag day$^{-1}$.
These features are unique among well-studied SN II-P. 

The color curves are illustrated in Fig. 6 and are compared to the curves
for SN 1999em. The $(B-V)$ curves of SN 2004A and 1999em are  
different, although the values of $(B-V)$ at early stage are similar.
The shape of $(V-R)$ and $(R-I)$ curves is similar for both objects.

As SN 2004ek exploded far from the center of its parent galaxy, 
at the distance close to the photometric radius, the extinction inside
the galaxy should be small. The Galactic extinction in the direction   
of UGC 724 is $A_V=0.22$ according to Schlegel et al. (1998). Accepting
this value as total extinction and the distance 73 Mpc from NED, we 
obtain maximum absolute magnitude $M_V=-17.95$, a high luminosity for
SN II-P. The absolute $V$-light curve for SN 2004ek is shown in 
Fig. 7. SN 2004ek is similar in luminosity to one of the brightest known
SN II-P 1992am (Schmidt et al., 1994), but the shape of the light curve
is different. The light curves of SNe 1999em and 2004ek can also be
compared on Fig. 7, and it is evident that the plateau lasted longer
for SN 2004ek and that the shape of the light curve on the plateau 
is different.

The reports on spectroscopic observations of SN 2004ek (Modjaz et al., 2004;
Filippenko et al., 2004) confirm the peculiar nature of this object.
Spectrum obtained on September 14.47 (about 6 days past explosion) showed
featureless blue continuum, and even on September 24 (16 days past 
outburst) the absorption lines were weak, and H$\alpha$ line was 
almost entirely in emission. The spectra of typical SNe II-P already show
prominent hydrogen Balmer lines with P-Cyg profiles at phase about
2-3 days, and at phase $\sim$16 days the absorption lines become much  
deeper. 

SN 2004ek has peculiar shape of the light curves in 
$BVRI$ bands and shape of $(B-V)$ color curve, long duration of the
plateau and high luminosity, unusual spectral evolution. Among
well-studied SNe II-P there is no analog to SN 2004ek. Unfortunately,
our observations of this object are not detailed, and we have no
data on the tail of the light curve. We appeal to all observers
who collected data on this SN to turn their attention to this 
interesting object, which may be important for revealing
the diversity of type II SNe. 

\medskip   

{\bf Acknowledgments:}
This research has made use of the NASA/IPAC Extragalactic Database
(NED) which is operated by the Jet Propulsion Laboratory, California
Institute of Technology, under contract with NASA.
The author is grateful to S.Yu.Shugarov, I.M.Volkov and N.N.Pavlyuk,
who made some observations, and to I.Myakishev and Meredith Rawls, 
who took part
in reductions of observations. The work was partly supported by the 
grant 05-02-17480 from RFBR. 

\references

Boles, T., Puckett, T., Cox, L., 2004, {\it IAU Circ.}, No. 8405

Filippenko, A.V., Ganeshalingam, M., Swift, B.J., 2004, 
{\it IAU Circ.}, No. 8411

Hamuy, M., Pinto, P.A., Maza, J., et al., 2001, 
{\it Astrophys. J.}, {\bf 558}, 615 

Hendry, M.A., Smartt, S.J., Crockett, R.M., et al., 2006,
{\it MNRAS}, {\bf 369}, 130
 
Elmhamdi, A., Danziger, I.J., Chugai, N., et al., 2003a,
{\it MNRAS}, {\bf 338}, 939 

Elmhamdi, A., Chugai, N.N., Danziger, I.J., 2003b,
{\it Astron. Astrophys.}, {\bf 404}, 1077
 
Kawakita, H., Kunigasa, K., Ayani, K., Yamaoka, H., 2004, 
{\it IAU Circ.}, No. 8266

Leonard, D.C., Filippenko, A.V., Gates, E.L., et al., 2002,
{\it PASP}, {\bf 114}, 35 

Modjaz, M., Kirshner, R., Challis, P., Matheson, T., 2004, 
{\it IAU Circ.}, No. 8409

Nakano, S., Itagaki, K., 2004, {\it IAU Circ.}, No. 8265

Schlegel, D., Finkbeiner, D., Davis, M., 1998, {\it Astrophys. J.},
{\bf 500}, 525 

Schmidt, B. P., Kirshner, R. P., Eastman, R. G., et al., 1994,
{\it Astron. J.}, {\bf 107}, 1444

Springob, C.M., Masters, K.L., Haynes, M.P., Giovanelli, R., 
Marinoni, C., 2007, {\it Astrophys. J. Suppl. Ser.}, {\bf 172},
599

Tsvetkov, D.Yu., Volnova, A.A., Shulga, A.P., Korotkiy, S.A., 
Elmhamdi, A., Danziger, I.J., Ereshko, M.V., 2006,
{\it Astron. Astrophys.}, {\bf 460}, 769

\endreferences

\PZfig{12cm}{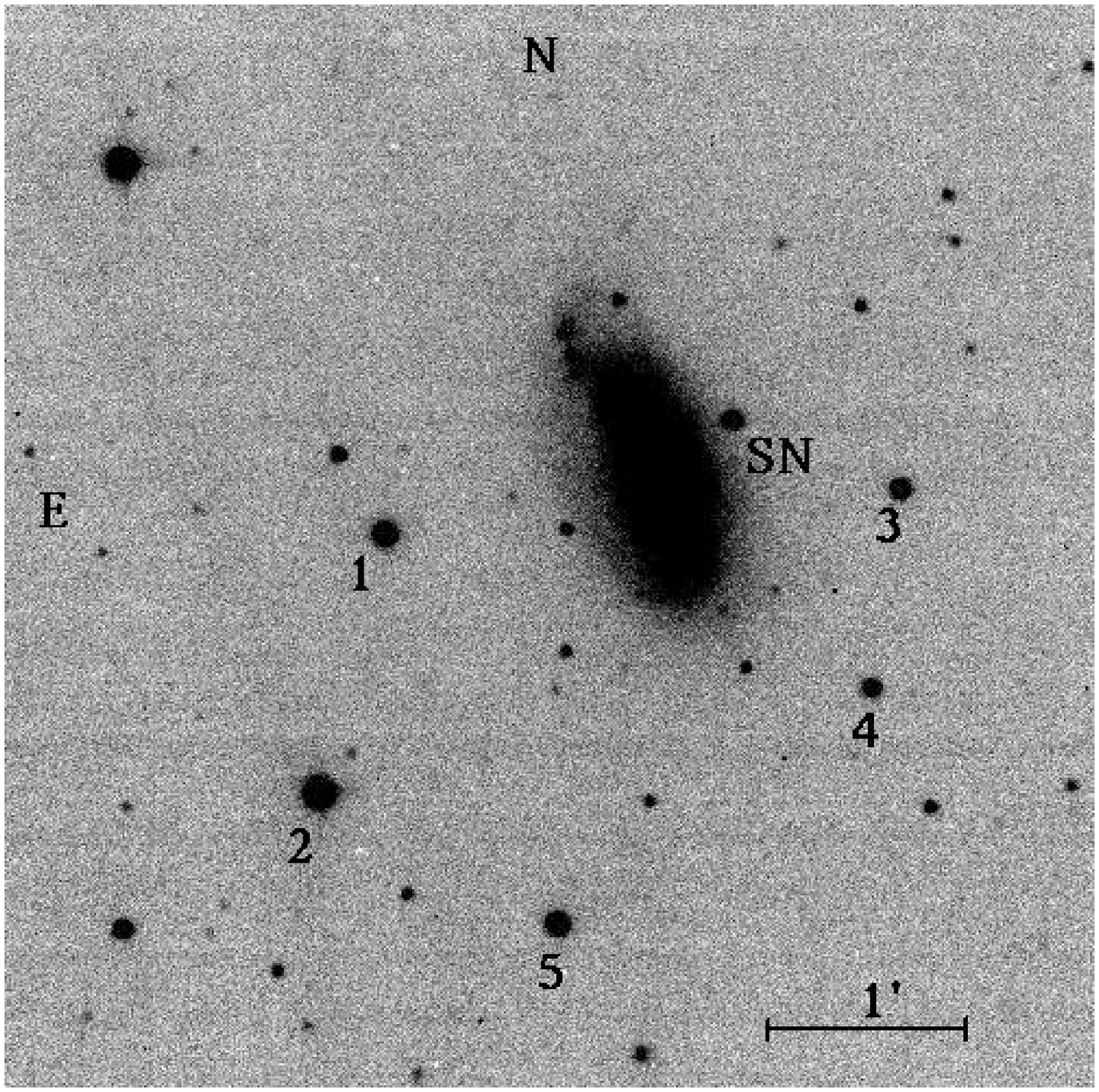}{SN 2004A in NGC 6207 with local standard
stars}

\PZfig{12cm}{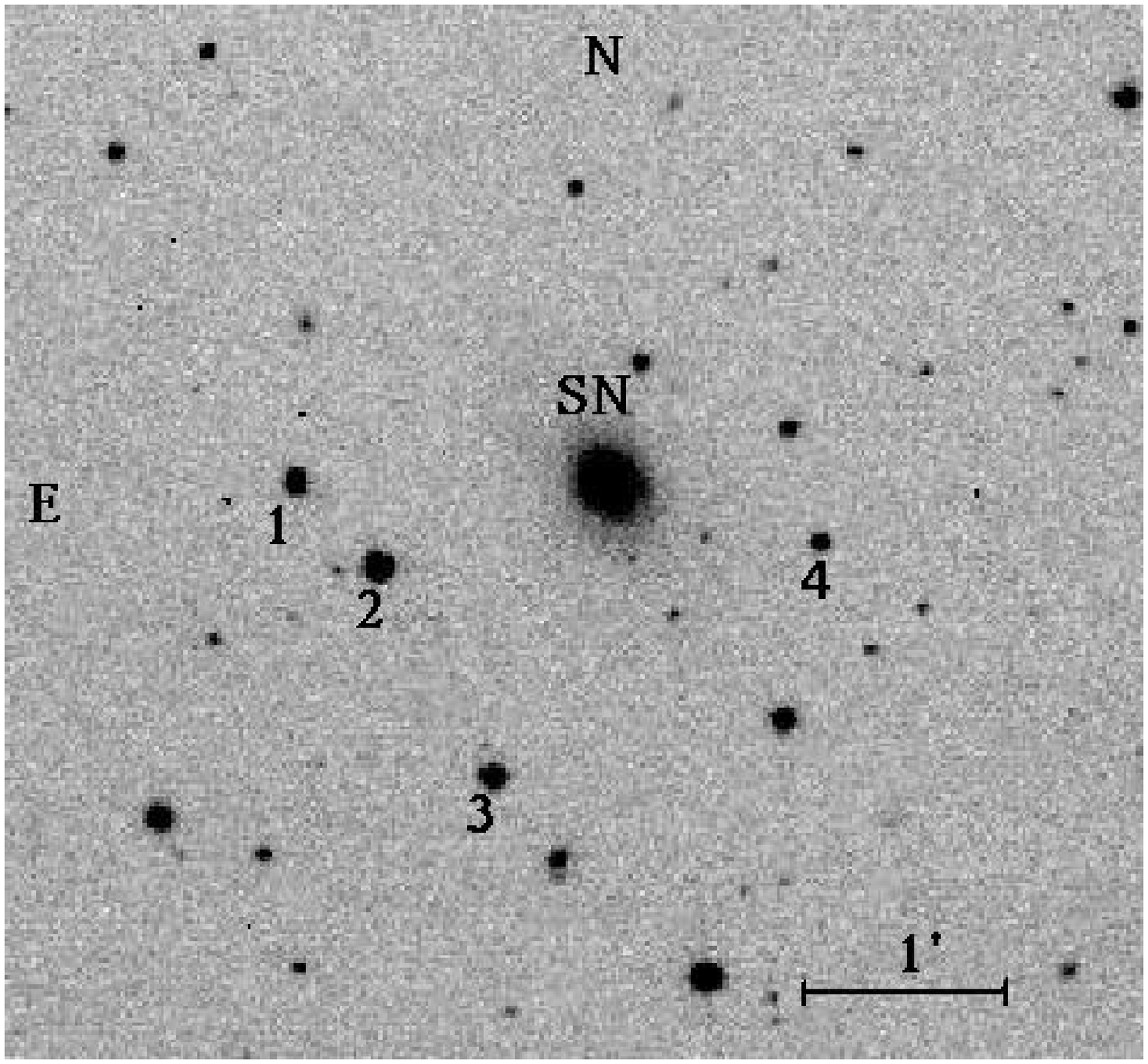}{SN 2004ek in UGC 724 with local standard
stars}

\PZfig{12cm}{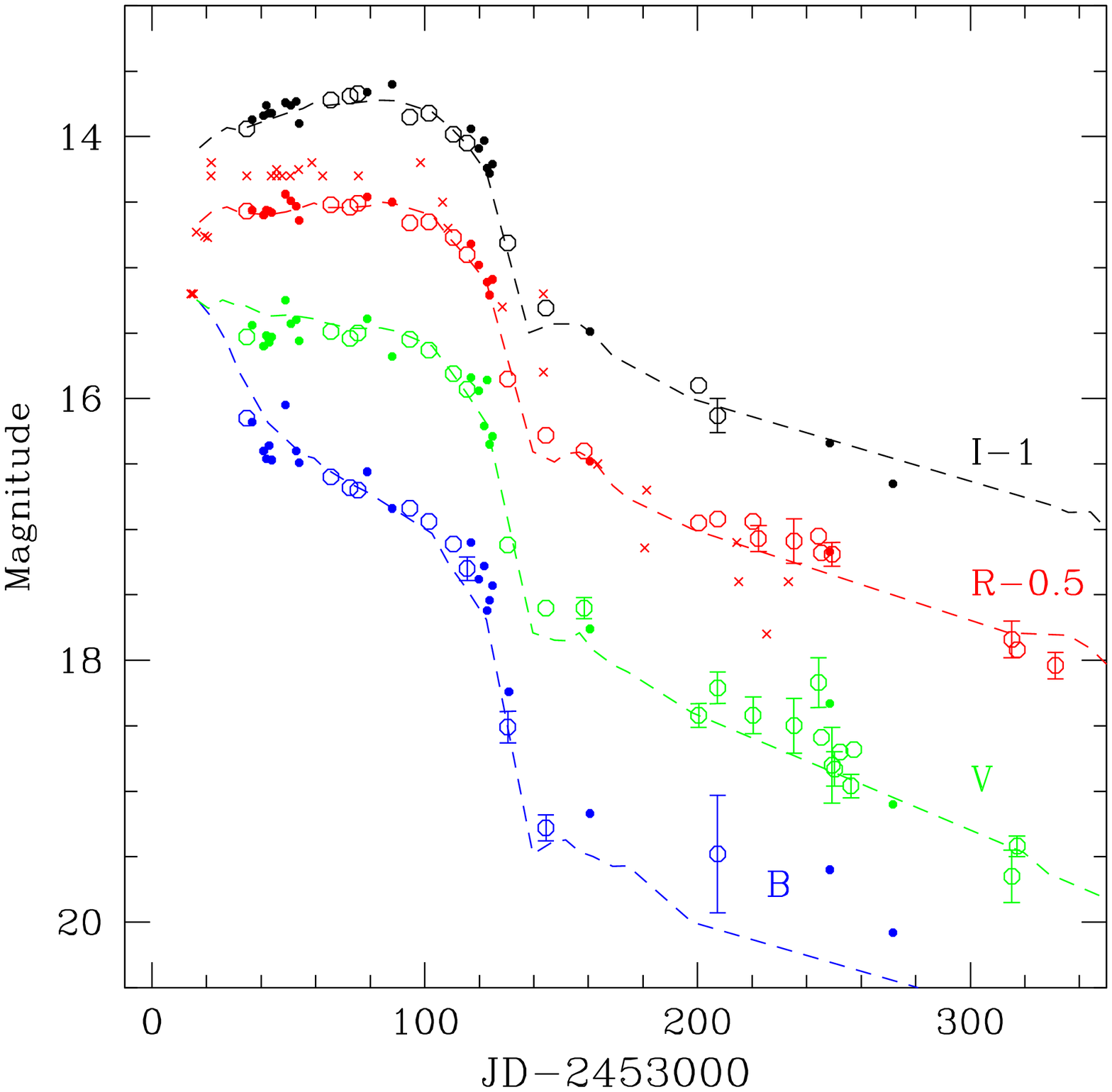}{$BVRI$ light curves of SN 2004A,
showing our photometry (circles), that of
Hendry et al. (2001) (dots) 
and the magnitudes reported at SNWeb (crosses). 
Error bars for our magnitudes are plotted only when they
exceed the size of a point. 
The dashed lines are the light curves of 
SN 1999em}

\PZfig{12cm}{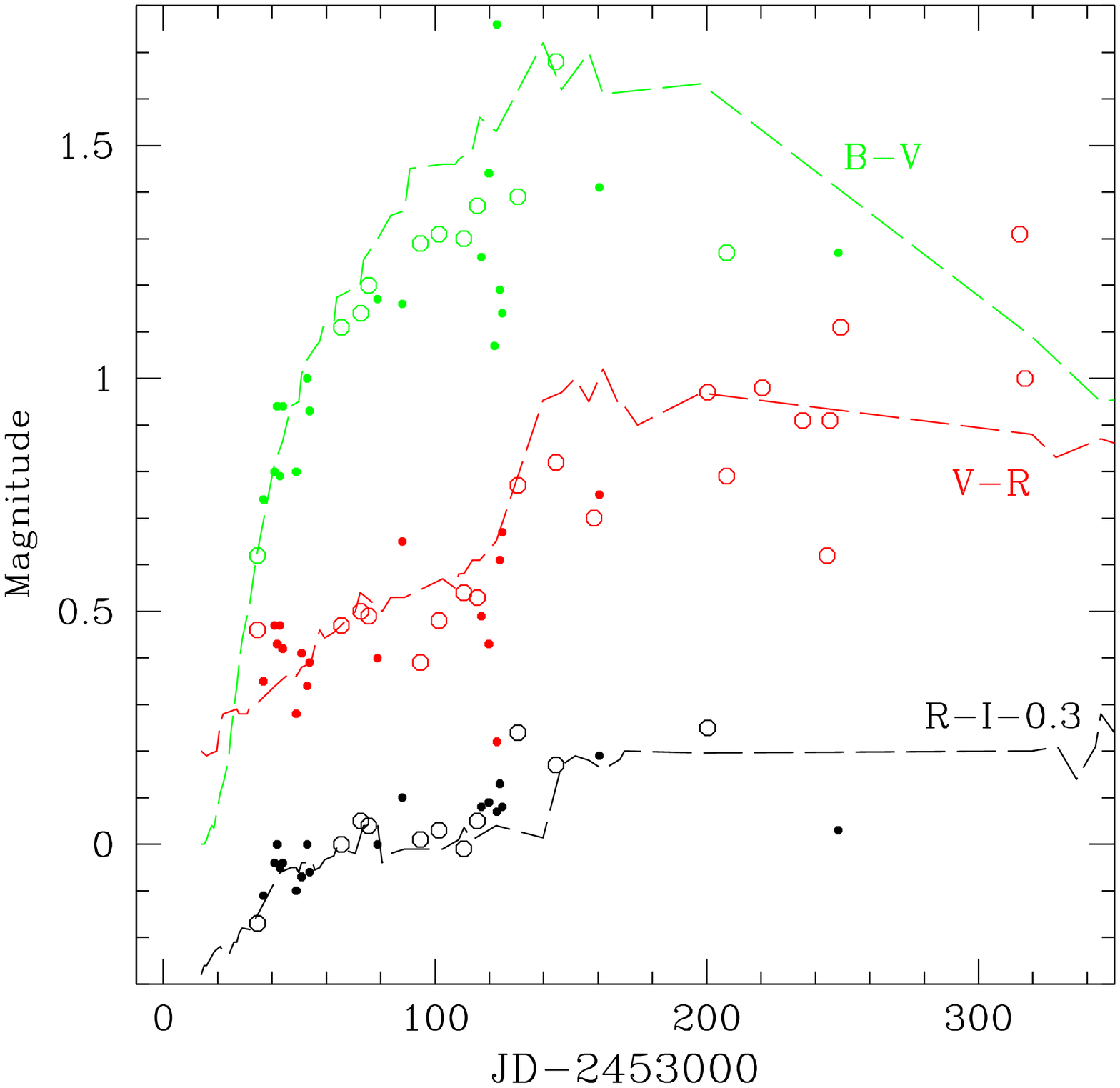}{The color curves of SN 2004A,
showing our photometry (circles) and that of   
Hendry et al. (2001) (dots). 
The dashed lines are the color curves of
SN 1999em}

\PZfig{12cm}{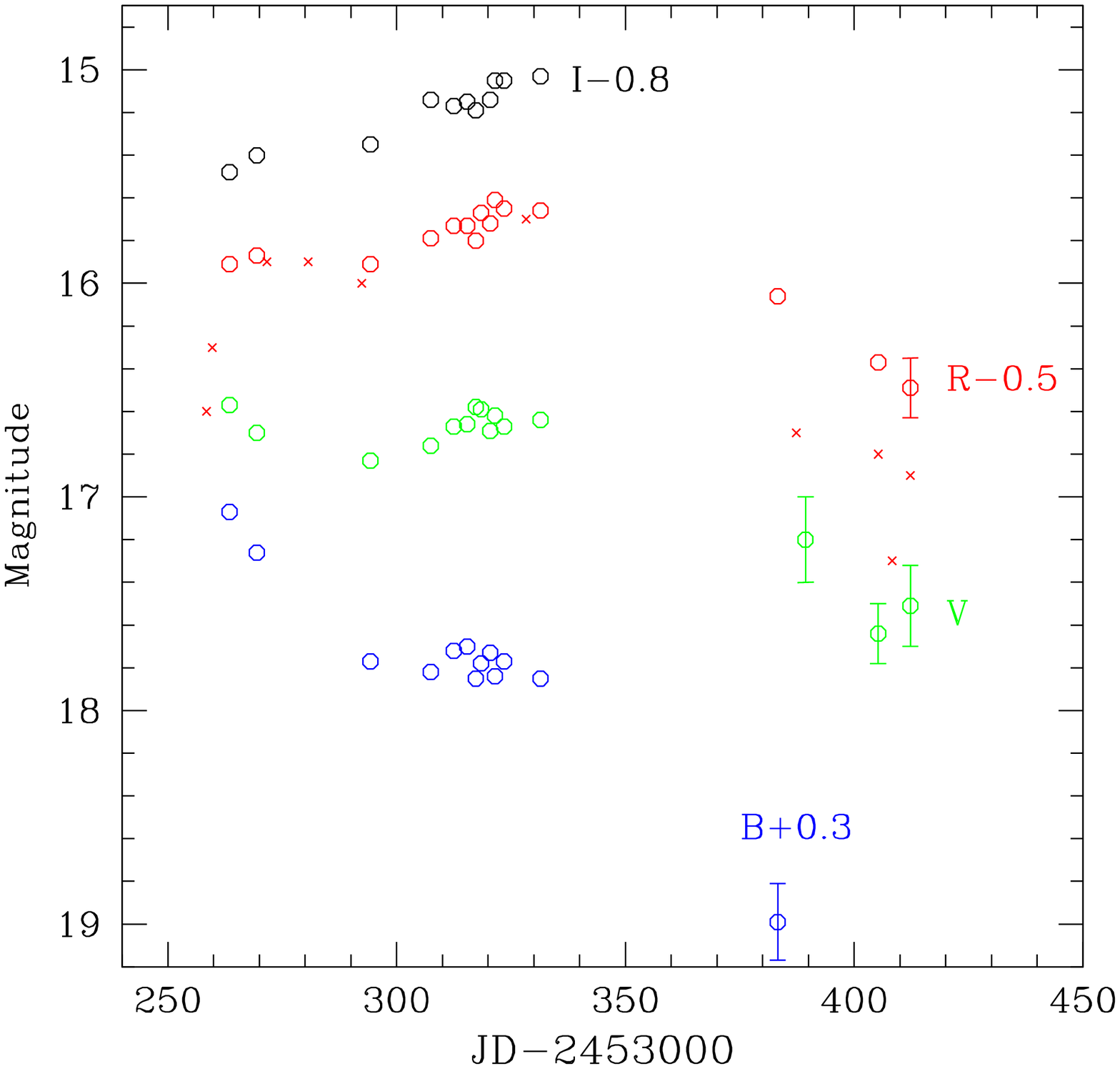}{$BVRI$ light curves of SN 2004ek.
Circles show
our data, crosses are for observations reported at SNWeb.
Error bars for our magnitudes are plotted only when they
exceed the size of a point}

\PZfig{12cm}{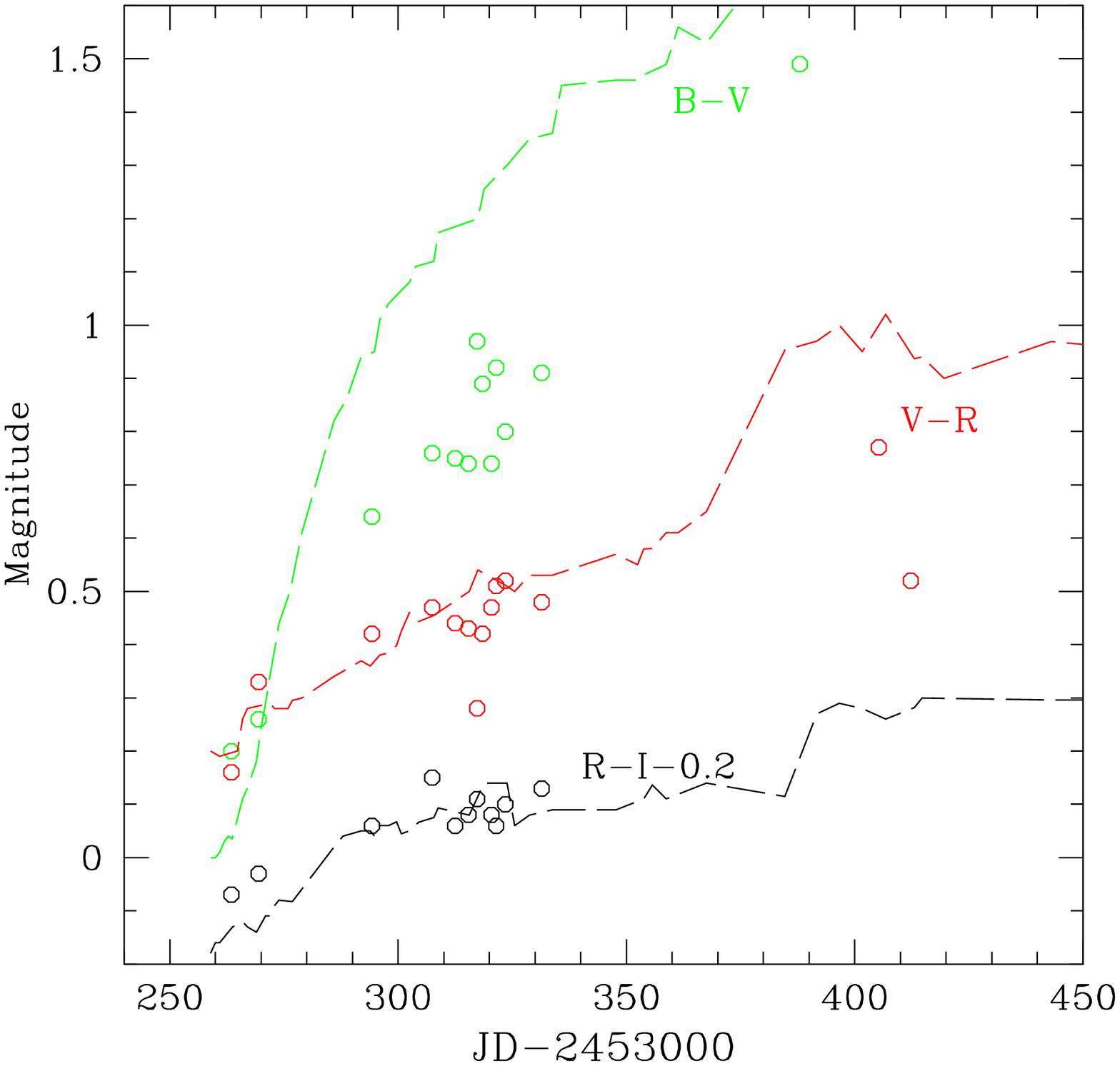}{The color curves of SN 2004ek.
The dashed lines are the color curves of
SN 1999em}

\PZfig{12cm}{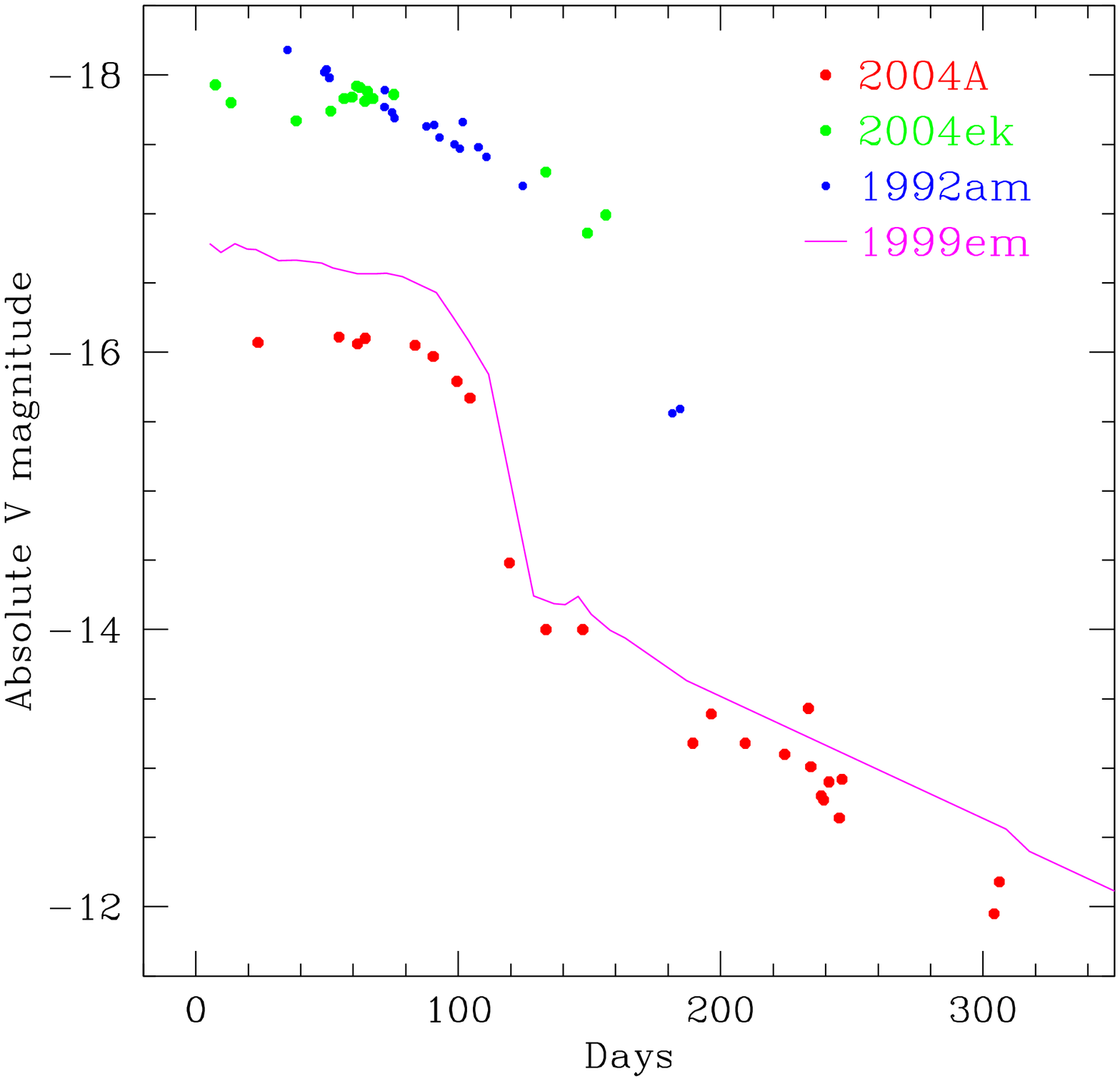}{The absolute $V$ light curves of SNe 2004A
and 2004ek compared to those for SNe 1999em and 1992am}

\newpage

\begin{table}
\caption{Magnitudes of local standard stars}\vskip2mm
\begin{tabular}{lcccccccccc}
\hline
Star &$U$ & $\sigma_U$ & $B$ & $\sigma_B$ & $V$ & $\sigma_V$ & $R$ &
$\sigma_R$ & $I$ & $\sigma_I$
\\
\hline
2004A-1  & 14.70& 0.07& 14.66& 0.04&  14.05& 0.01&  13.65& 0.03&  13.38& 0.03\\
2004A-2  & 14.55& 0.06& 14.00& 0.03&  13.11& 0.02&  12.55& 0.03&  12.15& 0.04\\
2004A-3  &      &     & 16.12& 0.03&  15.22& 0.01&  14.78& 0.02&  14.40& 0.04\\
2004A-4  &      &     & 16.39& 0.03&  15.66& 0.02&  15.27& 0.03&  14.94& 0.05\\
2004A-5  &      &     & 15.05& 0.04&  14.34& 0.02&  13.96& 0.03&  13.61& 0.05\\
2004ek-1 &      &     & 15.59& 0.01&  14.96& 0.01&  14.58& 0.01&  14.28& 0.01\\
2004ek-2 &      &     & 14.97& 0.01&  14.01& 0.01&  13.46& 0.01&  12.97& 0.02\\
2004ek-3 &      &     & 15.31& 0.01&  14.66& 0.01&  14.29& 0.02&  13.96& 0.02\\
2004ek-4 &      &     & 17.05& 0.04&  16.50& 0.02&  16.21& 0.03&  15.90& 0.05\\
\hline
\end{tabular}
\end{table}

\begin{table}
\caption{Observations of SN 2004A}\vskip2mm
\begin{tabular}{rccccccccl}
\hline
JD 2453000+ & $B$ & $\sigma_B$ & $V$ & $\sigma_V$ & $R$ & $\sigma_R$ &
$I$ & $\sigma_I$ & Tel.\\
\hline
  34.70&  16.15& 0.03&  15.53& 0.04&  15.07& 0.02&  14.94& 0.03& S50  \\
  65.53&  16.60& 0.02&  15.49& 0.02&  15.02& 0.02&  14.72& 0.02& C60b \\
  72.62&  16.68& 0.04&  15.54& 0.02&  15.04& 0.02&  14.69& 0.03& C60b \\
  75.56&  16.70& 0.03&  15.50& 0.02&  15.01& 0.02&  14.67& 0.03& M70a \\
  94.54&  16.84& 0.07&  15.55& 0.03&  15.16& 0.05&  14.85& 0.06& M70a \\
 101.48&  16.94& 0.03&  15.63& 0.02&  15.15& 0.01&  14.82& 0.02& M70a \\
 110.51&  17.11& 0.03&  15.81& 0.02&  15.27& 0.03&  14.98& 0.02& M70a \\
 115.52&  17.30& 0.09&  15.93& 0.03&  15.40& 0.03&  15.05& 0.04& M70a \\
 130.49&  18.51& 0.12&  17.12& 0.07&  16.35& 0.03&  15.81& 0.06& M70a \\
 144.44&  19.28& 0.10&  17.60& 0.06&  16.78& 0.02&  16.31& 0.04& M70a \\
 158.43&       &     &  17.60& 0.08&  16.90& 0.03&       &     & M70a \\
 200.40&       &     &  18.42& 0.09&  17.45& 0.05&  16.90& 0.07& M70b \\
 207.40&  19.48& 0.45&  18.21& 0.12&  17.42& 0.04&  17.13& 0.13& M70b \\
 220.39&       &     &  18.42& 0.14&  17.44& 0.06&       &     & M70b \\
 222.38&       &     &       &     &  17.57& 0.10&       &     & M70b \\
 235.38&       &     &  18.50& 0.21&  17.59& 0.17&       &     & M70b \\
 244.38&       &     &  18.17& 0.19&  17.55& 0.06&       &     & M70b \\
 245.37&       &     &  18.59& 0.06&  17.68& 0.03&       &     & C60a \\
 249.33&       &     &  18.80& 0.29&  17.69& 0.09&       &     & M70b \\
 250.26&       &     &  18.83& 0.13&       &     &       &     & C50  \\
 252.25&       &     &  18.70& 0.07&       &     &       &     & C50  \\
 256.26&       &     &  18.96& 0.09&       &     &       &     & C50  \\
 257.23&       &     &  18.68& 0.07&       &     &       &     & C50  \\
 315.21&       &     &  19.65& 0.20&  18.34& 0.14&       &     & C60a \\
 317.22&       &     &  19.42& 0.08&  18.42& 0.05&       &     & C60a \\
 331.24&       &     &       &     &  18.54& 0.10&       &     & C60a \\
\hline
\end{tabular}
\end{table}  

\newpage

\begin{table}
\caption{Observations of SN 2004ek}\vskip2mm
\begin{tabular}{rccccccccl}
\hline
JD 2453000+ & $B$ & $\sigma_B$ & $V$ & $\sigma_V$ & $R$ & $\sigma_R$ &
$I$ & $\sigma_I$ & Tel.\\
\hline
263.47&  16.77& 0.03&  16.57& 0.03 & 16.41& 0.02 & 16.28& 0.04& M70b \\
269.41&  16.96& 0.03&  16.70& 0.04 & 16.37& 0.04 & 16.20& 0.05& M70b \\
294.26&  17.47& 0.03&  16.83& 0.03 & 16.41& 0.03 & 16.15& 0.03& M70a \\
307.41&  17.52& 0.06&  16.76& 0.04 & 16.29& 0.03 & 15.94& 0.03& M70a \\
312.51&  17.42& 0.04&  16.67& 0.03 & 16.23& 0.03 & 15.97& 0.03& M70a \\
315.45&  17.40& 0.04&  16.66& 0.02 & 16.23& 0.02 & 15.95& 0.05& C60a \\
317.30&  17.55& 0.03&  16.58& 0.02 & 16.30& 0.02 & 15.99& 0.05& C60a \\
318.50&  17.48& 0.04&  16.59& 0.03 & 16.17& 0.04 &      &     & C60a \\
320.45&  17.43& 0.02&  16.69& 0.02 & 16.22& 0.02 & 15.94& 0.04& C60a \\
321.47&  17.54& 0.03&  16.62& 0.02 & 16.11& 0.01 & 15.85& 0.04& C60a \\
323.49&  17.47& 0.03&  16.67& 0.02 & 16.15& 0.02 & 15.85& 0.03& C60a \\
331.47&  17.55& 0.04&  16.64& 0.02 & 16.16& 0.02 & 15.83& 0.03& C60a \\
383.27&  18.69& 0.18&       &      & 16.56& 0.04 &      &     & M70a \\
389.37&       &     &  17.20& 0.20 &      &      &      &     & M70a \\
405.22&       &     &  17.64& 0.14 & 16.87& 0.07 &      &     & M70a \\
412.26&       &     &  17.51& 0.19 & 16.99& 0.14 &      &     & M70a \\
\hline
\end{tabular}
\end{table} 
\end{document}